\documentclass[12pt,thmsa]{article}

\begin{document}

\author{Axel Gelfert and Wolfgang Nolting \\
{\footnotesize Lehrstuhl Festk\"{o}rpertheorie, Institut f\"{u}r Physik }\\
{\footnotesize \ Humboldt-Universit\"{a}t zu Berlin, Invalidenstra\ss e 110 }%
\\
{\footnotesize \ 10115 Berlin, Germany}
\\
{\footnotesize \ email: gelfert@physik.hu-berlin.de}}
\title{Absence of a Magnetic Phase Transition in Heisenberg, Hubbard, and
Kondo-lattice (s-f) Films}
\date{}

\maketitle

\begin{abstract}
The possibility of a magnetic phase transition in Heisenberg, Hubbard, and
s-f (Kondo-lattice) films is investigated. It is shown that, for any finite
temperature $(\beta <\infty )$ and any finite number of layers $(d<\infty )$%
, the magnetization within every layer must vanish. Thus, the Mermin-Wagner
theorem is extended to a variety of system geometries. We also comment on
the microscopic interpretation of the transition from two to three
dimensions in the limit $d,\beta \rightarrow \infty $.
\end{abstract}

\section{Introduction}

In their well-known 1966 paper, Mermin and Wagner\cite{Mer Wag 1966} proved
rigorously that, at any finite temperature, there can be no spontaneous
magnetization in the one- and two-dimensional isotropic Heisenberg
ferromagnet. They were also able to exclude the possibility of spontaneous
sublattice magnetization in the one- and two-dimensional isotropic
two-sublattice antiferromagnet. The proof is based on an inequality
originally derived by Bogoliubov in a series of papers.\cite{Bog 1962} At
roughly the same time as Mermin and Wagner, Hohenberg showed there could be
no finite-temperature phase transition in one- and two-dimensional
superfluid systems.\cite{Hoh 1967} His proof, too, is based on Bogoliubov's
inequality.

Following these early papers, a lot of research has been carried out in
order to extend the Mermin-Wagner theorem to other systems. Wegner (1967)
considered a model describing a system with locally interacting itinerant
electrons.\cite{Weg 1967} Walker and Ruijgrok (1968) discussed a band model
for interacting electrons in a metal\cite{Wal Rui 1968}; Ghosh (1971), more
specifically, recovered the Mermin-Wagner theorem for the Hubbard
narrow-energy-band model.\cite{Gho 1971} A proof for the s-d interaction
model is given by van den Bergh and Vertogen (1974).\cite{Ber Ver 1974} A
paper by Robaszkiewicz and Micnas (1976) extends the Mermin-Wagner result to
a general model with localized and itinerant electrons, covering the
modified Zener model, the extended Hubbard model, s-d models, and the model
for the magnetic metal-insulator transition as particular cases.\cite{Rob
Mic 1976} In addition to extending the Mermin-Wagner theorem to several
microscopic many-body models, generalizations to more complicated geometries
were also found. E.g., Baryakhtar and Yablonskii (1975) prove the
Mermin-Wagner theorem for systems with an arbitrary number of magnetic
sublattices.\cite{Bar Yab 1975} Their proof not only excludes spontaneous
sublattice magnetization but also non-collinear magnetic order with an
external field being applied. Thorpe (1971) considers the case of
ferromagnetism in phenomenological models with double and higher-order
exchange terms\cite{Tho 1971}. These results were extended to the
multi-sublattice case by Krzemi\'{n}ski (1976).\cite{Krz 1976} Recently,
Matayoshi and Matayoshi (1997) have discussed models with $n$-th nearest
neighbour exchange interactions and tried to extend the Mermin-Wagner
theorem to anisotropic exchange interactions\cite{Mat Mat 1997}; their
proof, however, requires extremely special conditions on the parameters of
the model which appear to be of no physical significance.

The general procedure is essentially the same for most of the above papers:
The Bogoliubov inequality is used with suitable operators defined in such a
way as to give an upper bound for the desired order parameter, i.e. the
(bulk or sublattice) magnetization. In general, one arrives at an upper
bound that, in the limit $B_0\rightarrow 0$, behaves as $\sim (MB_0)^{1/3}$
or $\sim 1/(\ln MB_0)^{1/2}$ thus excluding a finite value of the order
parameter in one or two dimensions, resp. ($M$ denotes the bulk
magnetization and $B_0$ is the external magnetic field). In three
dimensions, no such behaviour is found, i.e. the usual line of reasoning
fails and spontaneous magnetization cannot be ruled out. Within the usual
scheme following Mermin and Wagner, the dimensionality of a system enters
into the calculation only through the volume element when integrating the
final version of the Bogoliubov inequality over $k$-space. Thus, a
microscopic picture of the transition from two to three dimensions does not
emerge.

Chester et al. (1969) argue\cite{Che et al 1969} that for three-dimensional
Bose systems of finite cross section or thickness the results of the
generically one or two-dimensional case are, in principle, reproduced. Their
main line of argument rests on the operators in the Bogoliubov inequality
being defined as Fourier transforms in a restricted space $D\times D\times L$
or $D\times L\times L,$ resp., where $D$ remains finite in the thermodynamic
limit while $L$ goes to infinity. This way, the operators lack the
physically intuitive meaning of the usual real-space or $k$-space operators,
which, again, means that no sound microscopic interpretation can be assigned
to the transition from two to three dimensions. Similar approaches have been
used by Fern\'{a}ndez (1970) and by Fisher and Jasnow (1971) to discuss systems of \textit{distinguishable} interacting quantum
particles and spin systems, respectively.\cite{Fer 1970b} \cite{Fis Jas 1971}

Today, as a result of the rapid progress of thin-film technology in recent
years, one can prepare and study systems with restricted geometries, such as
films, in great detail. Thus, important parameters, such as the Curie
temperature, magnetization and susceptibility, can be measured and discussed
as functions of the number of layers $d$ in a magnetic film. Experiments
indicate that, for real systems, the transition from 2D to 3D behaviour of,
for example, the critical exponent $\beta $, occurs within a narrow
crossover region of $d.$\cite{Li Bab 1991} The critical temperature $T_c$
also shows a strong $d$-dependence, with $T_c(d)$ quickly approaching the
bulk value $T_c(\infty )$ as $d$ increases (see \cite{Li Bab 1991},\cite{Far
1993}).

Taking these observations into account, it appears desirable to improve
one's theoretical understanding of how the transition from the monolayer to
the bulk occurs. By referring only to the truly microscopic properties of
the respective many-body model, it should be possible to study in detail the
transition from two to three dimensions. Work in this direction has been
done by studying, both analytically and numerically, film systems with
symmetry-breaking contributions to the Hamiltonian (for Heisenberg films
with single-ion anisotropy see \cite{Schi Nol 1999}), thus allowing the
study of the Curie temperature as a function of $d$ and the anisotropy
parameter. With respect to ferromagnetic itinerant-moment films,
layer/overlayer geometries have been investigated\cite{Wu Nol 1999}, giving
yet another indication of the relevance of many-body methods for film
systems.

At a more fundamental level, the validity of exact results, such as the
Mermin-Wagner theorem in two dimensions, to thin films may be tested. In
this paper, a proof of the Mermin-Wagner theorem for film systems within
three main many-body models (Heisenberg, Hubbard, and s-f model) will be
given.

For reasons of simplicity, we shall restrict our attention to film systems
composed of $d$ identical layers stacked on top of each other. The
calculations can easily be generalized to account for more complicated
geometries. One can think of the film geometry as consisting of a
two-dimensional Bravais lattice, i.e. the first layer, with $N$ lattice
sites and a $d$-atom basis that corresponds to the $d$ layers being stacked
up. Every lattice vector then decomposes into 
\begin{equation}
\vec{R}_{i\alpha }=\vec{R}_i+\vec{r}_\alpha
\end{equation}
where $\vec{R}_i$ is a vector of the Bravais lattice and $\vec{r}_\alpha $
is the basis vector pointing to the $\alpha $-th layer. In all of the
following calculations, Greek indices label layers and Roman indices refer
to sites of the Bravais lattice. It should be noted that translational
invariance can only be assumed within each layer. E.g., the notion of a 
\textit{reciprocal lattice} only makes sense when referring to the
two-dimensional Bravais lattice. A similar \textit{caveat} applies to
Fourier transforms being thought of as connecting real-space quantities with
those defined in wave-vector space.

\section{Model Hamiltonians}

In the following we shall discuss the Mermin-Wagner theorem for the
Heisenberg model, the Hubbard model, and the s-f model. For film systems one
has to distinguish between Bravais lattice indices and layer indices for all
site-dependent quantities, such as spin operators $S_{i\alpha
}^{(+,-,x,y,z)},$ annihilation and creation operators $c_{i\alpha }^{(+)}$
or coupling constants $J_{ij}^{\alpha \beta }$ which depend on two lattice
sites. In this notation, the Hamiltonian for Heisenberg films is given by

\begin{equation}
H=-\sum_{ij\alpha \beta }J_{ij}^{\alpha \beta }(S_{i\alpha }^{+}S_{j\beta
}^{-}+S_{i\alpha }^zS_{j\beta }^z)-b\sum_{i\alpha }e^{-i\vec{K}\cdot \vec{R}%
_i}S_{i\alpha }^z
\end{equation}
where the term $b\sum_{i\alpha }e^{-i\vec{K}\cdot \vec{R}_i}S_{i\alpha }^z$
is due to the interaction with an external magnetic field $b=\frac{g_J\mu
_BB_0}\hbar .$ The interaction with an external magnetic field leads to the
magnetization 
\begin{equation}
M=\frac 1{Nd}\frac{g_J\mu _B}\hbar \sum_{i\alpha }e^{-i\vec{K}\cdot \vec{R}%
_i}\left\langle S_{i\alpha }^z\right\rangle \equiv \frac 1d\sum_\alpha
M_\alpha
\end{equation}
where the phase factor $e^{-i\vec{K}\cdot \vec{R}_i}$ accounts for both
ferromagnetic and antiferromagnetic ordering, depending on the choice of $%
\vec{K}.$ We also assume the coupling constants $J_{ij}^{\alpha \beta }$ to
satisfy some general conditions, i.e. 
\begin{equation}
J_{ij}^{\alpha \beta }=J_{ji}^{\beta \alpha }
\end{equation}
\[
J_{ll}^{\varepsilon \varepsilon }=0 
\]
and 
\begin{equation}
\frac 1{Nd}\sum_{\gamma \varepsilon }\sum_{mp}\left| J_{pm}^{\varepsilon
\gamma }\right| \frac{\left( \vec{R}_m-\vec{R}_p\right) ^2}4\equiv \tilde{Q}%
<\infty .
\end{equation}
These conditions are very weak considering that for example the exchange
integrals $J$ will usually decay exponentially with distance.

As an example for itinerant-electron systems, we shall discuss the Hubbard
model

\begin{eqnarray}
H &=&\sum_{ij\alpha \beta \sigma }T_{ij}^{\alpha \beta }c_{i\alpha \sigma
}^{+}c_{j\beta \sigma }+\frac U2\sum_{i\alpha \sigma }n_{i\alpha \sigma
}n_{i\alpha -\sigma }-b\sum_{i\alpha }e^{-i\vec{K}\cdot \vec{R}_i}\sigma
_{i\alpha }^z  \nonumber \\
&=&\sum_{ij\alpha \beta \sigma }T_{ij}^{\alpha \beta }c_{i\alpha \sigma
}^{+}c_{j\beta \sigma }-\frac{2U}{3\hbar ^2}\sum_{i\alpha }\vec{\sigma}%
_{i\alpha }\cdot \vec{\sigma}_{i\alpha }+\frac{U\hat{N}_{tot}}%
2-b\sum_{i\alpha }e^{-i\vec{K}\cdot \vec{R}_i}\sigma _{i\alpha }^z
\label{Hubbard real-space}
\end{eqnarray}
where $T_{ij}^{\alpha \beta }$ describes the hopping of an electron from
lattice site $j$ of the $\beta $-th layer to site $i$ of the $\alpha $-th
layer, $U$ is the energy associated with having two electrons at the same
lattice site, and $\hat{N}_{tot}=\sum_{i\alpha }(n_{i\alpha \uparrow
}+n_{i\alpha \downarrow })$. Again, a term corresponding to an external
magnetic field is included. $\sigma _{i\alpha }^z$ in the Hubbard model is
the $z$-component of the spin of the electrons associated with the
respective lattice site, i.e. 
\begin{equation}
\sigma _{i\alpha }^z=\frac \hbar 2(c_{i\alpha \uparrow }^{+}c_{i\alpha
\uparrow }-c_{i\alpha \downarrow }^{+}c_{i\alpha \downarrow })
\end{equation}
Similar to the Heisenberg case, we require the hopping constants $%
T_{ij}^{\alpha \beta }$ to satisfy the isotropy conditions $T_{ij}^{\alpha
\beta }=T_{ji}^{\beta \alpha }$ as well as to converge upon summation over
all lattice sites: 
\begin{equation}
\frac 1{Nd}\sum_{\gamma \beta }\sum_{nk}\left| T_{nk}^{\gamma \beta }\right| 
\frac{\left( \vec{R}_n-\vec{R}_k\right) ^2}4\equiv \tilde{q}<\infty .
\end{equation}

In the s-f model one deals with two spin sub-systems, $\left\{ \vec{S}%
_{i\alpha }\right\} $ and $\left\{ \vec{\sigma}_{i\alpha }\right\} $, the
former consisting of localized $f$-electrons, the latter being associated
with itinerant $s$-electrons. With $z_{\sigma =_{\downarrow }^{\uparrow }}=\pm 1$, the s-f Hamiltonian is

\begin{equation}
H=\sum_{ij\alpha \beta \sigma }T_{ij}^{\alpha \beta }c_{i\alpha \sigma
}^{+}c_{j\beta \sigma }-\frac J2\sum_{i\alpha \sigma }(z_\sigma S_{i\alpha
}^zn_{i\alpha \sigma }+S_{i\alpha }^\sigma c_{i\alpha -\sigma
}^{+}c_{i\alpha \sigma })+\sum_{i\alpha }e^{-i\vec{K}\cdot \vec{R}%
_i}(S_{i\alpha }^z+\sigma _{i\alpha }^z).
\end{equation}

\section{Bogoliubov inequality and choice of operators}

The original Bogoliubov inequality as derived in \cite{Bog 1962} is 
\begin{equation}
\left| \left\langle \left[ C,A\right] _{-}\right\rangle \right| ^2\leq \frac
\beta 2\left\langle \left[ A,A^{+}\right] _{+}\right\rangle \left\langle
\left[ [C,H]_{-},C^{+}\right] _{-}\right\rangle
\end{equation}
where $A$ and $C$ are local operators and $\left\langle ....\right\rangle $
denotes the thermodynamic expectation value. It should be noted that the two
factors on the r.h.s. each are, mathematically, upper bounds to a norm and,
thus, greater than or equal to zero. In particular, if, for example, the
double commutator depends on some parameter $k,$ we will always have 
\begin{equation}
\left\langle \left[ \lbrack C,H]_{-},C^{+}\right] _{-}\right\rangle
(k)+\left\langle \left[ [C,H]_{-},C^{+}\right] _{-}\right\rangle (k^{\prime
})\geq \left\langle \left[ [C,H]_{-},C^{+}\right] _{-}\right\rangle (k)
\end{equation}

For our purposes a slightly modified version of the Bogoliubov inequality
shall be used. Dividing both sides by the double commutator and summing over
all wave vectors $\vec{k}$ associated with the two-dimensional Bravais
lattice, one arrives at 
\begin{equation}
\sum_{\vec{k}}\frac{\left| \left\langle \left[ C,A\right] _{-}\right\rangle
\right| ^2}{\left\langle \left[ [C,H]_{-},C^{+}\right] _{-}\right\rangle }%
\leq \frac \beta 2\sum_{\vec{k}}\left\langle \left[ A,A^{+}\right]
_{+}\right\rangle  \label{bog}
\end{equation}

The choice of suitable operators $A$ and $C$ is crucial; it determines
whether the inequality will be physically meaningful or not. In film systems
long-range magnetic order\textit{\ within a given layer} is conceivable
where the bulk (or even sub-lattice) magnetization of the whole system
vanishes, e.g. 
\begin{eqnarray*}
\ldots {} &\uparrow \uparrow \uparrow \uparrow \uparrow \uparrow \uparrow
\uparrow \uparrow \uparrow \uparrow \uparrow \uparrow \uparrow \uparrow
\uparrow \ldots \\
\ldots {} &\uparrow \downarrow \uparrow \downarrow \uparrow \downarrow
\uparrow \downarrow \uparrow \downarrow \uparrow \downarrow \uparrow
\downarrow \uparrow \downarrow \ldots \\
\ldots {} &\uparrow \downarrow \uparrow \downarrow \uparrow \downarrow
\uparrow \downarrow \uparrow \downarrow \uparrow \downarrow \uparrow
\downarrow \uparrow \downarrow \ldots \\
\ldots {} &\downarrow \downarrow \downarrow \downarrow \downarrow \downarrow
\downarrow \downarrow \downarrow \downarrow \downarrow \downarrow \downarrow
\downarrow \downarrow \downarrow \ldots
\end{eqnarray*}

Excluding long-range magnetic order for every layer within a film would,
therefore, be a considerably stronger statement. The general idea of our
proof is to use the Bogoliubov inequality to find an upper bound for the
layer magnetization $M_\beta ,$ i.e. 
\[
M_\beta \leq f(B_0,M) 
\]
where $f$ is a function that approaches zero as $B_0\rightarrow 0$ and does
not depend on any layer-specific quantities.

This is best achieved by choosing $C$ as a sum $\sum_\alpha (...)$ of spin
operators and $A$ as a spin operator associated with a specific layer $\beta
,$ say. This way, the numerator of the l.h.s of the Bogoliubov inequality
will be a layer-dependent quantity, while the double commutator in the
denominator will be summed over all lattice sites and thus be
layer-independent. One may then expect to be able to replace the r.h.s. of
the inequality by a (layer-independent) upper bound. More specifically, for
the \textit{Heisenberg model} we set 
\begin{equation}
A_{(\alpha )}\equiv S_\alpha ^{-}(-\vec{k}-\vec{K})
\end{equation}
and 
\begin{equation}
C\equiv \sum_\beta C_\beta \equiv \sum_\beta S_\beta ^{+}(\vec{k})
\end{equation}
where we have used the Fourier transform 
\begin{equation}
S_\alpha ^{(+,-,x,y,z)}(\vec{k})=\sum_ie^{i\vec{k}\cdot \vec{R}_i}S_{i\alpha
}^{(+,-,x,y,z)}
\end{equation}
Note that in $\vec{k}$-space we have $\left( S_\beta ^{+}(\vec{k})\right)
^{+}=S_\beta ^{-}(-\vec{k})$ due to the definition of the Fourier transform.
The operators $S$ in the context of the Heisenberg model are spin operators
for which the usual commutation relations hold true, such as, in $\vec{k}$%
-space, 
\begin{equation}
\left[ S_\alpha ^{+}(\vec{k}_1),S_\beta ^{-}(\vec{k}_2)\right] _{-}=2\hbar
\delta _{\alpha \beta }S_\alpha ^z(\vec{k}_1+\vec{k}_2)
\end{equation}
and 
\begin{equation}
\left[ S_\alpha ^z(\vec{k}_1),S_\beta ^{\pm }(\vec{k}_2)\right] _{-}=\pm
\hbar \delta _{\alpha \beta }S_\alpha ^{\pm }(\vec{k}_1+\vec{k}_2)
\end{equation}

In the\textit{\ Hubbard model }a similar definition for the operators $A$
and $C$ is used. However, as mentioned above, the operators $S^{(...)}$ now
are built up from fermionic creation and annihilation operators, e.g. 
\begin{equation}
\sigma _\alpha ^{-}(\vec{k}_1)=\hbar c_{\vec{k}_1\alpha \downarrow }^{+}c_{%
\vec{k}_1\alpha \uparrow }  \label{sigma minus}
\end{equation}
and 
\begin{equation}
\sigma _\beta ^{+}(\vec{k}_2)=\hbar c_{\vec{k}_2\uparrow \beta }^{+}c_{\vec{k%
}_2\beta \downarrow }  \label{sigma plus}
\end{equation}
The operators $A$ and $C$ are, then, 
\begin{equation}
A_{(\alpha )}\equiv \sigma _\alpha ^{-}(-\vec{k}-\vec{K})=\hbar c_{-\vec{k}-%
\vec{K},\alpha \downarrow }^{+}c_{-\vec{k}-\vec{K},\alpha \uparrow }
\end{equation}
and 
\begin{equation}
C\equiv \sum_\beta \sigma _\beta ^{+}(\vec{k})=\hbar \sum_\beta c_{\vec{k}%
\beta \uparrow }^{+}c_{\vec{k}\beta \downarrow }\left( =\sum_\beta C_\beta
\right) 
\end{equation}
respectively. The commutation relations for spin operators may, in a purely
formal sense, be used in this case as well.

In the \textit{s-f model}, one must be careful not to forget that one is
dealing with two separate spin sub-systems, one of which can be described by
usual spin operators, while the other is associated with itinerant
electrons. The two systems are independent from one another in the sense
that spin operators and creation or annihilation operators commute: 
\begin{equation}
\left[ S_{i\alpha }^{(+,-,x,y,z)},c_{j\beta \sigma }^{(+)}\right] _{-}=0
\end{equation}
and, thus, 
\begin{equation}
\left[ S_\alpha ^{(+,-,x,y,z)}(\vec{k}_1),c_{\vec{k}_2\beta \sigma
}^{(+)}\right] _{-}=0
\end{equation}

With (\ref{sigma plus}),(\ref{sigma minus}) we define as operators $A$ and $C
$ in the s-f model

\begin{equation}
A_{(\gamma )}=S_\gamma ^{-}(-\vec{k}-\vec{K})+\sigma _\gamma ^{-}(-\vec{k}-%
\vec{K})
\end{equation}
and 
\begin{equation}
C\equiv \sum_\beta C_\beta =\sum_\beta \left( S_\beta ^{+}(\vec{k})+\sigma
_\beta ^{+}(\vec{k})\right)
\end{equation}

\section{Evaluation of the Bogoliubov inequality}

\subsection{Hamiltonian-independent quantities}

It is obvious from the structure of the Bogoliubov inequality (\ref{bog})
that the numerators on both sides of the inequality are determined entirely
by the choice of operators as discussed in the previous section. The details
of the many-body model enter the calculation only via the double commutator $%
\left\langle \left[ [C,H]_{-},C^{+}\right] _{-}\right\rangle .$ For
simplicity, we shall, therefore, start with the quantity 
\begin{eqnarray}
&&\left\langle \left[ C,A\right] _{-}\right\rangle  \\
&=&\sum_\beta \left\langle \left[ S_\beta ^{+}(\vec{k})+\sigma _\beta ^{+}(%
\vec{k}),S_\gamma ^{-}(-\vec{k}-\vec{K})+\sigma _\gamma ^{-}(-\vec{k}-\vec{K}%
)\right] _{-}\right\rangle  \\
&=&\sum_{mn\beta }e^{i\vec{k}\cdot \vec{R}_m}e^{-i(\vec{k}+\vec{K})\cdot 
\vec{R}_n}\left\langle \left[ S_{m\beta }^{+}+\hbar c_{m\beta \uparrow
}^{+}c_{m\beta \downarrow },S_{n\gamma }^{-}+\hbar c_{n\gamma \downarrow
}^{+}c_{n\gamma \uparrow }\right] _{-}\right\rangle   \nonumber
\end{eqnarray}
for the s-f model. For reasons described above, $S$ and $\sigma $ operators
commute, so the commutator can be evaluated directly using the fundamental
commutation relations. This leads to

\begin{eqnarray}
\left\langle \left[ C,A\right] _{-}\right\rangle &=&2\hbar \sum_{mn\beta
}\delta _{mn}\delta _{\beta \gamma }e^{i\vec{k}\vec{R}_m}e^{-i(\vec{k}+\vec{K%
})\cdot \vec{R}_n}\left\langle S_{m\beta }^z+\sigma _{m\beta }^z\right\rangle
\nonumber \\
&=&\frac{2\hbar ^2N}{g_J\mu _B}M_\gamma (T,B_0)
\end{eqnarray}
where we have introduced the layer magnetization 
\begin{equation}
M_\gamma (T,B_0)=\frac 1N\frac{g_J\mu _B}\hbar \sum_ne^{-i\vec{K}\cdot \vec{R%
}_n}\left\langle S_{n\gamma }^z+\sigma _{n\gamma }^z\right\rangle
\end{equation}

It is obvious that in both the Heisenberg and the Hubbard model the same
relation 
\begin{equation}
\left\langle \left[ C,A_{(\gamma )}\right] _{-}\right\rangle =\frac{2\hbar
^2N}{g_J\mu _B}M_\gamma (T,B_0)
\end{equation}
holds, the only difference being the fact that only one spin system
contributes to the magnetization.

The r.h.s. of the Bogoliubov inequality is proportional to the
anticommutator sum 
\begin{equation}
\sum_{\vec{k}}\left\langle \left[ A,A^{+}\right] _{+}\right\rangle
\end{equation}
which, for the s-f model, is given by 
\begin{equation}
\sum_{\vec{k}}\left[ A,A^{+}\right] _{+}=\sum_{\vec{k}mn}e^{-i(\vec{k}+\vec{K%
})\cdot (\vec{R}_m-\vec{R}_n)}\left( \left[ \left( S_{n\gamma }^{-}+\hbar
c_{n\gamma \downarrow }^{+}c_{n\gamma \uparrow }\right) ,\left( S_{m\gamma
}^{+}+\hbar c_{m\gamma \uparrow }^{+}c_{m\gamma \downarrow }\right) \right]
_{+}\right)
\end{equation}
Summing the exponential over all $\vec{k}$ gives the delta function $N\delta
_{mn}.$ The mixed commutators involving both spin and creation/annihilation
operators in this case do not vanish, since we have the anticommutator
rather than the commutator. Thus, we arrive at the expression

\begin{eqnarray}
\left\langle \sum_{\vec{k}}\left[ A,A^{+}\right] _{+}\right\rangle
&=&N\sum_n\left( \left\langle \left[ S_{n\gamma }^{-},S_{n\gamma
}^{+}\right] _{+}\right\rangle +2\hbar \left\langle c_{n\gamma \downarrow
}^{+}c_{n\gamma \uparrow }S_{n\gamma }^{+}+c_{n\gamma \uparrow
}^{+}c_{n\gamma \downarrow }S_{n\gamma }^{-}\right\rangle \right.  \nonumber
\\
&&\left. +{\hbar }^2\left\langle c_{n\gamma \downarrow }^{+}c_{n\gamma
\uparrow }c_{n\gamma \uparrow }^{+}c_{n\gamma \downarrow }+c_{n\gamma
\uparrow }^{+}c_{n\gamma \downarrow }c_{n\gamma \downarrow }^{+}c_{n\gamma
\uparrow }\right\rangle \right)
\end{eqnarray}
\linebreak Since we are interested in the quantity $\left\langle \sum_{\vec{k%
}}\left[ A,A^{+}\right] _{+}\right\rangle $ as an upper bound in the
Bogoliubov inequality, it suffices to find upper bounds for the individual
expectation values on the r.h.s. For the\textit{\ Heisenberg term} we find

\begin{equation}
\sum_n\left\langle \left[ S_{n\gamma }^{-},S_{n\gamma }^{+}\right]
_{+}\right\rangle =2\sum_n\left\langle (S_{n\gamma }^x)^2+(S_{n\gamma
}^y)^2\right\rangle \leq 2\sum_n\left\langle \vec{S}_{n\gamma
}^2\right\rangle =2\hbar ^2S(S+1)N
\end{equation}
For the \textit{Hubbard contribution} we have 
\begin{eqnarray}
\sum_n\left\langle c_{n\gamma \downarrow }^{+}c_{n\gamma \uparrow
}c_{n\gamma \uparrow }^{+}c_{n\gamma \downarrow }+c_{n\gamma \uparrow
}^{+}c_{n\gamma \downarrow }c_{n\gamma \downarrow }^{+}c_{n\gamma \uparrow
}\right\rangle &\leq &\sum_n\left\langle c_{n\gamma \downarrow
}^{+}c_{n\gamma \downarrow }\left( 1-c_{n\gamma \uparrow }^{+}c_{n\gamma
\uparrow }\right) \right.  \nonumber \\
&&\left. +c_{n\gamma \uparrow }^{+}c_{n\gamma \uparrow }\left( 1-c_{n\gamma
\downarrow }^{+}c_{n\gamma \downarrow }\right) \right\rangle  \nonumber \\
&\leq &\sum_n\left( \left\langle n_{n\gamma \downarrow }\right\rangle
+\left\langle n_{n\gamma \uparrow }\right\rangle \right)  \nonumber \\
&\leq &2N
\end{eqnarray}

For the \textit{mixed terms} appearing only in the s-f model, an upper bound
is given by

\begin{equation}
\sum_{n\sigma }\left\langle c_{n\gamma \sigma }^{+}c_{n\gamma -\sigma
}S_{n\gamma }^{-\sigma }\right\rangle \leq 2\left( 4+2S(S+1)\right) N
\end{equation}
as shown in appendix A (see eqn. (\ref{gemischte c-s})).

Tabulating these results we have 
\[
\sum_{\vec{k}}\left\langle \left[ A,A^{+}\right] _{+}\right\rangle \leq
2\hbar ^2S(S+1)N^2 
\]
for the Heisenberg model; 
\[
\sum_{\vec{k}}\left\langle \left[ A,A^{+}\right] _{+}\right\rangle \leq
2\hbar ^2N^2 
\]
for the Hubbard model; and 
\[
\sum_{\vec{k}}\left\langle \left[ A,A^{+}\right] _{+}\right\rangle \leq
N^2\hbar ^2\left( 4S(S+1)+10\right) 
\]
for the s-f model.

\subsection{The double commutator $\left\langle \left[
[C,H]_{-},C^{+}\right] _{-}\right\rangle $}

In this section, we shall calculate the remaining double commutator $\left\langle \left[ [C,H]_{-},C^{+}\right]
_{-}\right\rangle $ for the individual models and give upper bounds for use in the Bogoliubov inequality.

\subsubsection{The Heisenberg case}

The double commutator $\left\langle \left[ [C,H]_{-},C^{+}\right]
_{-}\right\rangle $ in this case is 
\begin{eqnarray}
\left\langle \left[ \lbrack C,H]_{-},C^{+}\right] _{-}\right\rangle 
&=&\sum_{\gamma \varepsilon }\left\langle \left[ \left[ S_\gamma ^{+}(\vec{k}%
),H_0+H_b\right] _{-},S_\varepsilon ^{-}(-\vec{k})\right] _{-}\right\rangle 
\nonumber \\
&=&\sum_{\gamma \varepsilon }\sum_{mp}e^{-i\vec{k}\cdot (\vec{R}_m-\vec{R}%
_p)}\left\langle \left[ [S_{m\gamma }^{+},H]_{-},S_{p\varepsilon
}^{-}\right] _{-}\right\rangle   \label{Heisen double commut}
\end{eqnarray}
The real-space commutator on the r.h.s. can be easily evaluated using the
standard commutation relations. Thus, one arrives at 
\begin{eqnarray}
\left\langle \left[ \lbrack C,H]_{-},C^{+}\right] _{-}\right\rangle (\vec{k}%
) &=&\sum_{\gamma \varepsilon }\sum_{mp}J_{pm}^{\varepsilon \gamma }\hbar
^2\left( \left( 1-e^{-i\vec{k}\cdot (\vec{R}_m-\vec{R}_p)}\right) \cdot
\right.   \nonumber \\
&&\left. \cdot \left\langle 2S_{p\varepsilon }^zS_{m\gamma }^z+S_{m\gamma
}^{+}S_{p\varepsilon }^{-}\right\rangle \right)  \\
&&+2b\hbar ^2\sum_\varepsilon \sum_me^{-i\vec{K}\cdot \vec{R}_m}\left\langle
S_{m\varepsilon }^z\right\rangle   \nonumber
\end{eqnarray}
To this we add the double commutator $\left\langle \left[
[C,H]_{-},C^{+}\right] _{-}\right\rangle (-\vec{k})$ which, as discussed
above, is a positive real number. Replacing the spin operator expectation
values by the upper bound $2\hbar ^2S^2,$ we find, after some minor algebra, 
\begin{equation}
\left\langle \left[ \lbrack C,H]_{-},C^{+}\right] _{-}\right\rangle \leq
4Nd\hbar ^2\left| B_0M(T,B_0)\right| +8\hbar ^4S^2\sum_{\gamma \varepsilon
}\sum_{mp}\left| J_{pm}^{\varepsilon \gamma }\right| \frac{\vec{k}^2\cdot (%
\vec{R}_m-\vec{R}_p)^2}4
\end{equation}
where we have already used the fact that 
\begin{equation}
1-\cos (\vec{k}\cdot (\vec{R}_m-\vec{R}_p))\leq \frac{\vec{k}^2\cdot (\vec{R}%
_m-\vec{R}_p)^2}4.
\end{equation}
With the above definition of the constant $\tilde{Q},$ we arrive at the
final result 
\begin{equation}
\left\langle \left[ \lbrack C,H]_{-},C^{+}\right] _{-}\right\rangle \leq
4Nd\hbar ^2\left( \left| B_0M(T,B_0)\right| +2\hbar ^2S^2\tilde{Q}\vec{k}%
^2\right) 
\end{equation}

\subsubsection{The Hubbard case}

Again, we need to calculate the full double commutator $\left\langle \left[
[C,H]_{-},C^{+}\right] _{-}\right\rangle $. With the Hubbard Hamiltonian
given in eqn. (\ref{Hubbard real-space}), we have
\begin{equation}
\sum_\gamma \left[ \sigma _\gamma ^{+}(\vec{k}),H\right] _{-}=\hbar
\sum_{ij\alpha \beta }T_{ij}^{\alpha \beta }c_{i\alpha \uparrow
}^{+}c_{j\beta \downarrow }\left( e^{-i\vec{k}\cdot \vec{R}_i}-e^{-i\vec{k}%
\cdot \vec{R}_j}\right) +b\hbar \sum_\alpha \sigma _\alpha ^{+}(\vec{k}+\vec{%
K})
\end{equation}
which still needs to be commuted with $\sum_\varepsilon \sigma _\varepsilon
^{-}(-\vec{k})$ .  Replacing the expectation values by their modulus, one gets the relevant inequality  
\begin{eqnarray}
\left\langle \left[ \lbrack C,H]_{-},C^{+}\right] _{-}\right\rangle (\vec{k}%
) &=&\sum_{\gamma \varepsilon }\left\langle \left[ \left[ \sigma _\gamma
^{+}(\vec{k}),H_0+H_b\right] _{-},\sigma _\varepsilon ^{-}(-\vec{k})\right]
_{-}\right\rangle   \nonumber \\
&\leq &\hbar ^2\sum_{il\alpha \varepsilon }T_{il}^{\alpha \varepsilon
}\left( e^{-i\vec{k}\cdot (\vec{R}_i-\vec{R}_l)}-1\right) \left( \left|
\left\langle c_{i\alpha \uparrow }^{+}c_{l\varepsilon \uparrow
}\right\rangle \right| +\left| \left\langle c_{l\varepsilon \downarrow
}^{+}c_{i\alpha \downarrow }\right\rangle \right| \right)   \nonumber \\
&&+2b\hbar ^2\sum_\alpha \left\langle \sigma _\alpha ^z(\vec{K}%
)\right\rangle .
\end{eqnarray}
For the same reasons as above, we may now add $\left\langle \left[
[C,H]_{-},C^{+}\right] _{-}\right\rangle (-\vec{k})$ to get an upper bound
for the l.h.s. For the expectation values $\left| \left\langle c_{i\alpha
\uparrow }^{+}c_{l\varepsilon \uparrow }\right\rangle \right| +\left|
\left\langle c_{l\varepsilon \downarrow }^{+}c_{i\alpha \downarrow
}\right\rangle \right| $ we find as an upper bound $\left| \left\langle
c_{i\alpha \uparrow }^{+}c_{l\varepsilon \uparrow }\right\rangle \right|
+\left| \left\langle c_{l\varepsilon \downarrow }^{+}c_{i\alpha \downarrow
}\right\rangle \right| \leq 4$ (see appendix A, eqn. (\ref{gemischte c}),
for detailed calculation), so in total we may write, using the above
notation, 
\begin{equation}
\left\langle \left[ \lbrack C,H]_{-},C^{+}\right] _{-}\right\rangle \leq
4Nd\hbar ^2\left( \left| B_0M(T,B_0)\right| +2\tilde{q}k^2\right) .
\end{equation}
\subsubsection{The s-f case}

The s-f double commutator 
\begin{equation}
\left\langle \left[ \lbrack C,H]_{-},C^{+}\right] _{-}\right\rangle
=\sum_{\gamma \varepsilon }\left\langle \left[ \left[ S_\gamma ^{+}(\vec{k}%
)+\sigma _\gamma ^{+}(\vec{k}),H_0+H_b\right] _{-},S_\varepsilon ^{-}(-\vec{k%
})+\sigma _\varepsilon ^{-}(-\vec{k})\right] _{-}\right\rangle 
\end{equation}
may be computed by considering first the field-independent contribution to
the Hamiltonian 
\begin{equation}
H_0=\sum_{ij\alpha \beta \sigma }T_{ij}^{\alpha \beta }c_{i\alpha \sigma
}^{+}c_{j\beta \sigma }-\frac J2\sum_{i\alpha \sigma }\left( z_\sigma
S_{i\alpha }^zn_{i\alpha \sigma }+S_{i\alpha }^\sigma c_{i\alpha -\sigma
}^{+}c_{i\alpha \sigma }\right) 
\end{equation}
and the respective commutators  
\begin{equation}
\left[ \sigma _{m\beta }^{+},H_0\right] _{-}=\sum_{k\gamma }T_{km}^{\gamma
\beta }\hbar \left( c_{m\beta \uparrow }^{+}c_{m\beta \downarrow
}-c_{k\gamma \uparrow }^{+}c_{m\beta \downarrow }\right) -J\hbar \left(
S_{m\beta }^{+}\sigma _{m\beta }^z-2S_{m\beta }^z\sigma _{m\beta
}^{+}\right) 
\end{equation}
\begin{equation}
\left[ S_{m\beta }^{+},H_0\right] _{-}=J\hbar \left( S_{m\beta }^{+}\sigma
_{m\beta }^z-2S_{m\beta }^z\sigma _{m\beta }^{+}\right) 
\end{equation}
We then find (neglecting, for the moment, the external contribution $H_b$)  
\begin{eqnarray}
\left\langle \left[ \lbrack C,H_0]_{-},C^{+}\right] _{-}\right\rangle (\vec{k%
}) &=&\sum_{kmn}\sum_{\beta \gamma \delta }e^{-i\vec{k}\cdot (\vec{R}_m-\vec{%
R}_n)}T_{km}^{\gamma \beta }\hbar ^2\cdot   \nonumber \\
&&\cdot \left\langle \left[ \left( c_{m\beta \uparrow }^{+}c_{m\beta
\downarrow }-c_{k\gamma \uparrow }^{+}c_{m\beta \downarrow }\right) ,\left(
S_{n\delta }^{-}+c_{n\delta \downarrow }^{+}c_{n\delta \uparrow }\right)
\right] _{-}\right\rangle   \nonumber \\
&=&-\sum_{kn}\sum_{\beta \gamma \delta }\left( 1-e^{i\vec{k}\cdot (\vec{R}_n-%
\vec{R}_m)}\right) \cdot   \nonumber \\
&&\cdot \left( \delta _{\delta \beta }T_{kn}^{\gamma \beta }c_{n\delta
\downarrow }^{+}c_{k\gamma \downarrow }+\delta _{\delta \gamma
}T_{nk}^{\gamma \beta }c_{k\beta \uparrow }^{+}c_{n\delta \uparrow }\right)
\hbar ^2
\end{eqnarray}
Following the usual procedure of adding $\left\langle \left[
[C,H]_{-},C^{+}\right] _{-}\right\rangle (-\vec{k})$ we arrive at the upper
bound 
\begin{eqnarray}
\left\langle \left[ \lbrack C,H_0]_{-},C^{+}\right] _{-}\right\rangle  &\leq
&2\hbar ^2\sum_{nk\gamma \beta }\left| T_{nk}^{\gamma \beta }\right| \left(
1-\cos \left( \vec{k}\cdot \left( \vec{R}_n-\vec{R}_k\right) \right) \right)
\cdot   \nonumber \\
&&\cdot \left( \left| \left\langle c_{n\beta \downarrow }^{+}c_{k\gamma
\downarrow }\right\rangle \right| +\left| \left\langle c_{n\beta \uparrow
}^{+}c_{k\gamma \uparrow }\right\rangle \right| \right)   \nonumber \\
&\leq &2\hbar ^2\sum_{nk\gamma \beta }\left| T_{nk}^{\gamma \beta }\right| 
\vec{k}^2\left( \vec{R}_n-\vec{R}_k\right) ^2  \nonumber \\
&=&2Nd\tilde{q}\hbar ^2\vec{k}^2
\end{eqnarray}
where $\left| \left\langle c_{n\beta \downarrow }^{+}c_{k\gamma \downarrow
}\right\rangle \right| +\left| \left\langle c_{n\beta \uparrow
}^{+}c_{k\gamma \uparrow }\right\rangle \right| \leq 4$ has already been
used.

To this preliminary result, we still need to add the double commutator $%
\left\langle \left[ [C,H_b]_{-},C^{+}\right] _{-}\right\rangle $ with the
external part $H_b$ of the Hamiltonian. Here, too, we need to take into
account the presence of two distinct spin systems. Thus, $H_b$ is given by 
\begin{equation}
H_b=-B_0\frac{\mu _B}\hbar \sum_{i\alpha }(g_JS_{i\alpha }^z+2\sigma
_{i\alpha }^z)e^{-i\vec{K}\cdot \vec{R}_i}
\end{equation}
where the usual definitions of the spin operators apply. As in the previous
cases, one gets 
\begin{equation}
\left\langle \left[ \lbrack C,H_b]_{-},C^{+}\right] _{-}\right\rangle
=-2\hbar ^2NdB_0M(T,B_0)
\end{equation}
where, however, the magnetization is now defined as 
\begin{equation}
M(T,B_0)=\frac 1{Nd}\frac{\mu _B}\hbar \sum_{i\beta }e^{-i\vec{K}\cdot \vec{R%
}_i}\left( \left\langle g_JS_{i\beta }^z\right\rangle +2\left\langle \sigma
_{i\beta }^z\right\rangle \right) 
\end{equation}

Finally, we arrive at 
\begin{equation}
\left\langle \left[ \lbrack C,H]_{-},C^{+}\right] _{-}\right\rangle \leq
2Nd\hbar ^2\left( \left| B_0M(T,B_0)\right| +\tilde{q}\vec{k}^2\right)
\end{equation}

\section{Proving the absence of spontaneous magnetization}

Within all three models, we have found for the double commutator 
\begin{equation}
\left\langle \left[ \lbrack C,H]_{-},C^{+}\right] _{-}\right\rangle \leq \xi
_0^2Nd\left( \left| B_0M(T,B_0)\right| +\xi _1\vec{k}^2\right)
\end{equation}
where $\xi _i$ are constants depending, at most, on fixed parameters of the
respective many-body models. We also know that 
\begin{equation}
\left\langle \left[ C,A\right] _{-}\right\rangle =\xi _2NM_\gamma (T,B_0)
\end{equation}
and 
\begin{equation}
\sum_{\vec{k}}\left\langle \left[ A,A^{+}\right] _{+}\right\rangle \leq 2\xi
_3N^2
\end{equation}

We can now give a generally applicable discussion of the layer
magnetization. With the Bogoliubov inequality (\ref{bog}),

\begin{equation}
\sum_{\vec{k}}\frac{\left| \left\langle \left[ C,A\right] _{-}\right\rangle
\right| ^2}{\left\langle \left[ [C,H]_{-},C^{+}\right] _{-}\right\rangle }%
\leq \frac \beta 2\sum_{\vec{k}}\left\langle \left[ A,A^{+}\right]
_{+}\right\rangle
\end{equation}
we find 
\begin{equation}
\sum_{\vec{k}}\frac{\xi _2^2N^2M_\gamma ^2(T,B_0)}{\xi _0^2Nd\left( \left|
B_0M(T,B_0)\right| +\xi _1\vec{k}^2\right) }\leq \xi _3\beta N^2
\end{equation}
The l.h.s. of the inequality can be replaced by an integral, using the
formula 
\begin{equation}
\sum_{\vec{k}}=\frac{L^2}{(2\pi )^2}\int\limits_{\vec{k}}d^2\vec{k},
\end{equation}
where $\frac{L^2}{(2\pi )^2}$ is the area in two-dimensional $\vec{k}$-space
associated with one quantum state. Restricting the support of the integral
to a finite-volume sphere inscribed into the first Brillouin zone only
strengthens the inequality, so 
\begin{equation}
\left( \frac{\xi _2}{\xi _0}\right) ^2\frac 1{2\pi d}\frac{L^2}NM_\gamma
^2(T,B_0)\int\limits_ 0^{k_0}dk\frac k{\left| B_0M(T,B_0)\right| +\xi
_1k^2}\leq \xi _3\beta
\end{equation}
where $k_0$ is the cutoff corresponding to the sphere in $\vec{k}$-space. In
the thermodynamic limit, $\frac{L^2}N=v_0^{(2)}$ approaches a constant
finite value as $N\rightarrow \infty .$ Evaluating the integral and
performing some minor algebra, we have 
\begin{equation}
M_\gamma ^2(T,B_0)\leq \xi \frac{\beta d}{\ln \left( 1+\frac{\xi _1k_0^2}{%
\left| B_0M(T,B_0)\right| }\right) }
\end{equation}
($\xi $ again is an unsignificant constant).

From this formula, it is obvious that, for any finite temperature $(\beta
<\infty )$ and finite thickness $(d<\infty ),$ the logarithm in the
denominator will diverge in the limit $B_0\rightarrow 0,$ thus forcing the
layer magnetization $M_\gamma $ to vanish. One should note that the final
result does not depend on the choice of $\vec{K}$ and, thus, excludes both
ferromagnetic and antiferromagnetic ordering. This proves the Mermin-Wagner
theorem for film systems within the Heisenberg, Hubbard, and the s-f
(Kondo-lattice) model.

\section{Summary}

In this paper, we have been able to prove that the Mermin-Wagner theorem
which was originally shown to exclude a magnetic phase transition at finite
temperature for one and two-dimensional systems can be extended to
Heisenberg, Hubbard, and s-f (Kondo-lattice) films of \textit{any finite
thickness}. While this may be expected from very general considerations, and
similar theorems for Bose systems and essentially classical systems were
proposed before, a microscopic calculation for many-fermion systems has been
lacking so far.

In view of recent progress in the experimental and theoretical study of thin
magnetic films, it appears to be worthwhile to reconsider the importance of
the Mermin-Wagner theorem. This is particularly the case when the
symmetry-breaking is not due to explicitly adding anisotropic contributions
to the Hamiltonian, but is instead a secondary result of using numerical
approximations that violate the Mermin-Wagner theorem. Future work in this
direction might be able to provide useful insights for theorists and
experimentalists alike.

Our microscopic approach suggests that the parameter $d$, i.e. the number of
layers in a film, plays a similar role as the inverse temperature $\beta .$
If either one diverges, a phase transition cannot be ruled out. It is, of
course, not \textit{a priori }clear if the behaviour of an upper bound on a
physical quantity has any physical significance itself; however, we believe
our results to indicate that, within the discussed models, in order to
possibly describe (anti)ferromagnetism, one has to take the thermodynamic
limit seriously, in the sense that $N\rightarrow \infty $ does not suffice,
but rather $d\rightarrow \infty $ is required as well.

\section*{Acknowledgments}

Fruitful discussions with Dr. Michael Potthoff are gratefully acknowledged.
This work was supported by the Deutsche Forschungsgemeinschaft (DFG) within
the SFB 290.

\appendix 

\section{Upper bound for $\sum_\sigma \left\langle c_{n\gamma \sigma
}^{+}c_{n\gamma -\sigma }S_{n\gamma }^{-\sigma }\right\rangle $ and $\left|
\left\langle c_{n\beta \sigma }^{+}c_{k\gamma \sigma }\right\rangle \right| $%
}

We first discuss the quantity $\sum_\sigma \left\langle c_{n\gamma \sigma
}^{+}c_{n\gamma -\sigma }S_{n\gamma }^{-\sigma }\right\rangle $. The problem
in evaluating the expectation value arises from the fact, that we are
dealing with seemingly unrelated operators. It is, therefore, convenient to
use the identity 
\begin{eqnarray}
c_{n\gamma \sigma }^{+}c_{n\gamma -\sigma }S_{n\gamma }^{-\sigma } &=&\frac
14\left\{ \left( c_{n\gamma \sigma }^{+}c_{n\gamma -\sigma }+S_{n\gamma
}^\sigma \right) \left( c_{n\gamma -\sigma }^{+}c_{n\gamma \sigma
}+S_{n\gamma }^{-\sigma }\right) \right.  \nonumber \\
&&-\left( c_{n\gamma \sigma }^{+}c_{n\gamma -\sigma }-S_{n\gamma }^\sigma
\right) \left( c_{n\gamma -\sigma }^{+}c_{n\gamma \sigma }-S_{n\gamma
}^{-\sigma }\right)  \nonumber \\
&&+i\left( c_{n\gamma \sigma }^{+}c_{n\gamma -\sigma }+iS_{n\gamma }^\sigma
\right) \left( c_{n\gamma -\sigma }^{+}c_{n\gamma \sigma }-iS_{n\gamma
}^{-\sigma }\right)  \nonumber \\
&&\left. -i\left( c_{n\gamma \sigma }^{+}c_{n\gamma -\sigma }-iS_{n\gamma
}^\sigma \right) \left( c_{n\gamma -\sigma }^{+}c_{n\gamma \sigma
}+iS_{n\gamma }^{-\sigma }\right) \right\}  \nonumber \\
&\equiv &\frac 14\sum_{j=1}^4\phi (j)B_{j\sigma }B_{j\sigma }^{+}
\end{eqnarray}
where $j$ labels the individual terms and $\phi $ is a phase factor ($\phi
(1)=+1,$ $\phi (2)=-1,$ $\phi (3)=i,$ $\phi (4)=-i).$ The original
expectation value has thus been decomposed into the sum of expectation
values of pairs of adjunct operators $B_{j\sigma }B_{j\sigma }^{+}.$ Using
the spectral theorem, we find 
\begin{equation}
\sum_\sigma \left\langle c_{n\gamma \sigma }^{+}c_{n\gamma -\sigma
}S_{n\gamma }^{-\sigma }\right\rangle =\frac 1{4\hbar
}\sum_{j=1}^4\int\limits_{-\infty }^\infty dE\frac 1{e^{\beta E}+1}\phi
(j)\sum_\sigma S_{B_{j\sigma }^{+}B_{j\sigma }}^{(-)}(E)
\end{equation}
where $S_{B_{j\sigma }^{+}B_{j\sigma }}^{(-)}(E)=\frac 1{2\pi }\left\langle
\left[ B_{j\sigma }^{+},B_{j\sigma }\right] _{+}\right\rangle (E)$ is the
spectral density in its energy representation. The l.h.s. is the expectation
value of the sum of two adjunct operators and, thus, is real. We may now
make use of the fact that for pairs of adjunct operators, the spectral
density is positive definite, which, together with the triangle inequality
and $\left| \phi (j)\right| =1$ gives an upper bound for the r.h.s. : 
\begin{equation}
\frac 1{4\hbar }\sum_{j=1}^4\int\limits_{-\infty }^\infty dE\frac 1{e^{\beta
E}+1}\phi (j)\sum_\sigma S_{B_{j\sigma }^{+}B_{j\sigma }}^{(-)}(E)\leq \frac
14\sum_\sigma \sum_{j=1}^4\frac 1\hbar \int\limits_{-\infty }^\infty
dES_{B_{j\sigma }^{+}B_{j\sigma }}^{(-)}(E)
\end{equation}
The sum on the r.h.s. now consists of the $0$-th spectral moments associated
with the operator pairs $B_{j\sigma }^{+},B_{j\sigma }.$ Each of the
spectral moments is given by the expectation value of the anticommutator 
\begin{equation}
\frac 1\hbar \int\limits_{-\infty }^\infty dES_{B_{j\sigma }^{+}B_{j\sigma
}}^{(-)}(E)\equiv M_{B_{j\sigma }^{+}B_{j\sigma }}^{(0)}=\left\langle \left[
B_{j\sigma }^{+},B_{j\sigma }\right] _{+}\right\rangle .
\end{equation}
The anticommutators can be easily evaluated: 
\begin{eqnarray}
M_{B_{1\sigma }^{+}B_{1\sigma }}^{(0)} &=&\left\langle \left[ c_{n\gamma
-\sigma }^{+}c_{n\gamma \sigma }+S_{n\gamma }^{-\sigma },c_{n\gamma \sigma
}^{+}c_{n\gamma -\sigma }+S_{n\gamma }^\sigma \right] _{+}\right\rangle 
\nonumber \\
&\leq &\left\langle n_{n\gamma \sigma }-2n_{n\gamma \sigma }n_{n\gamma
-\sigma }+n_{n\gamma -\sigma }\right\rangle +2S(S+1)  \nonumber \\
&&+2\left\langle S_{n\gamma }^\sigma c_{n\gamma -\sigma }^{+}c_{n\gamma
\sigma }+S_{n\gamma }^{-\sigma }c_{n\gamma \sigma }^{+}c_{n\gamma -\sigma
}\right\rangle  \nonumber \\
&\leq &4+2S(S+1)+2\left\langle S_{n\gamma }^\sigma c_{n\gamma -\sigma
}^{+}c_{n\gamma \sigma }+S_{n\gamma }^{-\sigma }c_{n\gamma \sigma
}^{+}c_{n\gamma -\sigma }\right\rangle \\
M_{B_{2\sigma }^{+}B_{2\sigma }}^{(0)} &\leq &4+2S(S+1)-2\left\langle
S_{n\gamma }^\sigma c_{n\gamma -\sigma }^{+}c_{n\gamma \sigma }+S_{n\gamma
}^{-\sigma }c_{n\gamma \sigma }^{+}c_{n\gamma -\sigma }\right\rangle \\
M_{B_{3\sigma }^{+}B_{3\sigma }}^{(0)} &\leq &4+2S(S+1)+2i\left\langle
S_{n\gamma }^\sigma c_{n\gamma -\sigma }^{+}c_{n\gamma \sigma }-S_{n\gamma
}^{-\sigma }c_{n\gamma \sigma }^{+}c_{n\gamma -\sigma }\right\rangle \\
M_{B_{4\sigma }^{+}B_{4\sigma }}^{(0)} &\leq &4+2S(S+1)-2i\left\langle
S_{n\gamma }^\sigma c_{n\gamma -\sigma }^{+}c_{n\gamma \sigma }-S_{n\gamma
}^{-\sigma }c_{n\gamma \sigma }^{+}c_{n\gamma -\sigma }\right\rangle
\end{eqnarray}
and hence 
\begin{equation}
\sum_\sigma \left\langle c_{n\gamma \sigma }^{+}c_{n\gamma -\sigma
}S_{n\gamma }^{-\sigma }\right\rangle \leq 2\left( 4+2S(S+1)\right)
\label{gemischte c-s}
\end{equation}

For the expectation value $\left| \left\langle c_{n\beta \sigma
}^{+}c_{k\gamma \sigma }\right\rangle \right| $ one uses the same procedure.
For $(n,\beta )=(k,\gamma )$ it is obvious that $\left\langle c_{n\beta
\sigma }^{+}c_{n\beta \sigma }\right\rangle \leq 1.$ Starting with the
decomposition 
\begin{eqnarray}
c_{i\alpha \sigma }^{+}c_{l\nu \sigma } &=&\frac 14\left\{ \left( c_{i\alpha
\sigma }^{+}+c_{l\nu \sigma }^{+}\right) \left( c_{i\alpha \sigma }+c_{l\nu
\sigma }\right) \right.  \nonumber \\
&&-\left( c_{i\alpha \sigma }^{+}-c_{l\nu \sigma }^{+}\right) \left(
c_{i\alpha \sigma }-c_{l\nu \sigma }\right)  \nonumber \\
&&+i\left( c_{i\alpha \sigma }^{+}+ic_{l\nu \sigma }^{+}\right) \left(
c_{i\alpha \sigma }-ic_{l\nu \sigma }\right)  \nonumber \\
&&\left. -i\left( c_{i\alpha \sigma }^{+}-ic_{l\nu \sigma }^{+}\right)
\left( c_{i\alpha \sigma }+ic_{l\nu \sigma }\right) \right\}
\end{eqnarray}
one can write the expectation value as 
\begin{eqnarray}
\left| \left\langle c_{n\beta \sigma }^{+}c_{k\gamma \sigma }\right\rangle
\right| &=&\frac 14\sum_{j=1}^4\frac 1\hbar \int\limits_{-\infty }^\infty
dE\left| \phi (j)\right| \frac 1{e^{\beta E}+1}S_{A_jA_j^{+}}^{(-)}(E) 
\nonumber \\
&\leq &\frac 14\sum_{j=1}^4M_{A_jA_j^{+}}^{(0)}
\end{eqnarray}
(similar notation as above).

Since we have discussed the case $(n,\beta )=(k,\gamma )$ separately, we can
now assume $(n,\beta )\neq (k,\gamma ).$ The $0$-th spectral moments can be
easily calculated; they each give $M_{A_jA_j^{+}}^{(0)}=2.$ Thus for all $%
n,k,\beta ,\gamma $ we have as an upper bound 
\begin{equation}
\left| \left\langle c_{n\beta \sigma }^{+}c_{k\gamma \sigma }\right\rangle
\right| \leq 2.  \label{gemischte c}
\end{equation}


\begin{thebibliography}{99}
\bibitem{Mer Wag 1966}  Mermin, N.D. and H.\ Wagner: Phys. Rev. Lett. 17
(1966) 1133

\bibitem{Bog 1962}  Bogoliubov, N.N.: Phys. Abhandl. Sowjetunion 6 (1962),
1, 113, 229

\bibitem{Hoh 1967}  Hohenberg, P.C.: Phys. Rev. 158 (1967) 383

\bibitem{Weg 1967}  Wegner, F.: Phys. Lett. 24A (1967) 131

\bibitem{Wal Rui 1968}  Walker, M.B. and T.W. Ruijgrok: Phys. Rev. 171
(1968) 170

\bibitem{Gho 1971}  Ghosh, D.K.: Phys. Rev. Lett. 27 (1971) 1584

\bibitem{Ber Ver 1974}  van den Bergh, M. and G. Vertogen: Phys. Lett. A 50
(1974) 85

\bibitem{Rob Mic 1976}  Robaszkiewicz, S. and R. Micnas: phys. stat. sol.
(b) 73 (1976) K35

\bibitem{Bar Yab 1975}  Baryakhtar, V.G. and D.A. Yablonskii: phys. stat.
sol. (b) 72 (1975) K95

\bibitem{Tho 1971}  Thorpe, M.F.: J. appl. Phys. 42 (1971) 1410

\bibitem{Krz 1976}  Krzemi\'{n}ski, S.: phys. stat. sol. (b) 74 (1976) K119

\bibitem{Mat Mat 1997}  Matayoshi, G. and S. Matayoshi: Bull. Coll. Sci.
(Univ. Ryukyus) 64 (1997) 1

\bibitem{Che et al 1969}  Chester, G.V.; Fisher, M.E.; and N.D. Mermin:
Phys. Rev. 185 (1969) 760

\bibitem{Fer 1970b}  Fern\'{a}ndez, J.F.: Phys. Rev. A 2 (1970) 2555

\bibitem{Fis Jas 1971}  Fisher, M.D. and D. Jasnow: Phys. Rev. B 3 (1971) 907

\bibitem{Li Bab 1991}  Li, Y. and K. Baberschke: Phys. Rev. Lett. 68 (1992)
1208

\bibitem{Far 1993}  Farle, M.; Baberschke, K.; Stetter, U.; Aspelmeier, A.;
and F. Gerhardter: Phys. Rev. B 47 (1993) 11 572

\bibitem{Schi Nol 1999}  Schiller, R. and W. Nolting: Sol. St. Commun. 110
(1999) 121

\bibitem{Wu Nol 1999}  Wu, J.H.; Herrmann, T.; and W. Nolting: Phys. Rev. B
(in press)

\end{thebibliography}
\end{document}